\documentstyle[multicol,aps,epsfig]{revtex}
\draft
\begin{document}

\newcommand{\xminus}{\mbox{$X^-$}}
\newcommand{\xplus}{\mbox{$X^+$}}
\newcommand{\hminus}{\mbox{$H^-$}}
\newcommand{\dminus}{\mbox{$D^-$}}

\title{ Internal Transitions of Two-Dimensional Charged
Magneto-Excitons $X^-$: \\ Theory and Experiment}
\author{A. B. Dzyubenko\cite{ABD}}
\address{
 Institut f\"{u}r Theoretische Physik, J.W. Goethe-Universit\"{a}t,
 60054 Frankfurt,  Germany }
\author{A. Yu. Sivachenko}
\address{The Weizmann Institute of Science, Rehovot 76100, Israel}
\author{H. A. Nickel, T. M. Yeo, G. Kioseoglou, B. D. McCombe, and A. Petrou}
\address{Department of Physics and Center for Advanced Photonic and
Electronic Materials, SUNY Buffalo, Buffalo, NY 14260, USA}
\date{\today}
\maketitle
%
\begin{abstract}
Internal
spin-singlet and spin-triplet transitions of 
charged excitons $X^-$ in magnetic fields in quantum wells have been
studied experimentally and theoretically.
The allowed $X^-$ transitions are photoionizing and exhibit a characteristic
double-peak structure, which reflects the rich structure of the
magnetoexciton continua in higher Landau levels (LL's).
We discuss a novel exact selection rule,
a hidden manifestation of translational invariance, that governs
transitions of charged mobile complexes
in a magnetic field.
\end{abstract}
%

\pacs{73.20.Dx, 71.70.Di, 76.40.+b, 78.90.+t}

\begin{multicols}{2}

%
Recently there has been considerable interest in charged excitonic
complexes, $X^-$ and $X^+$, commonly
referred to as trions. The $X^-$ complex, which can be observed in
photoluminescence- and reflectance spectra of low-density,
quasi-2D electron gases (2DEGs), has been the subject of extensive
experimental and theoretical work since its observation in 1993
\cite{Kheng}.
The bulk of this work to date has been concerned
with {\em inter-\/}band transitions only. {\em Intra-\/}band, or
internal, transitions of $X^-$, which lie in the far-infrared (FIR),
 can provide additional important insight into the
properties of the ground and excited states of this complex.

The $X^-$-complex, consisting of an exciton binding an additional
electron, is superficially similar to its close relative, the
negatively charged donor ion, $D^-$ \cite{jiang}.
Both complexes are often considered to be the
semiconductor analogs of the hydrogen ion, $H^-$.
When the $H^-$ ion is treated in the infinite
proton mass approximation, such analogy is exact for the
$D^-$ complex --- a localized positive charge binding two electrons.
This analogy fails, however, in certain very important aspects for the
{\em mobile} $X^-$ complex.
In particular, we show here that the magnetic translations
for the $X^-$ imply the existence of an exact selection rule
that {\em prohibits} certain bound-to-bound internal $X^-$ transitions,
the analogs of which are very strong for the $D^-$.
In an arbitrary uniform $B$ this selection rule is applicable
to charged electron-hole, as well as to one-component electron systems.
In the latter case Kohn's theorem \cite{Kohn} based on translational
invariance also works. Due to the center-of-mass separation for
electron systems in $B$,
both theorems --- though based on different operator algebras ---
give in this case equivalent predictions.
%
%
%
To understand the main qualitative features, we first
consider the strictly-2D electron-hole ($e$--$h$)
system in high magnetic fields.
In this limit,
$\hbar\omega_{ce}, \hbar\omega_{ch} \gg
E_0 = \sqrt{\pi/2} \, e^2/\epsilon l_B$,
where
$E_0$ is the binding energy of the 2D magnetoexciton (MX)
in zero LL's \cite{L&L80} and $l_B =  (\hbar c/e B)^{1/2}$.
The mixing between different LL's can then be neglected,
and the $X^-$ states can be classified
by total electron and hole LL numbers, $(N_eN_h)$.
The corresponding basis for $X^-$ is of the form \cite{Dz_PLA}
$\phi^{(e)}_{n_1 m_1}({\bf r}) \,
 \phi^{(e)}_{n_2 m_2}({\bf R}) \,
 \phi^{(h)}_{N_{h}M_{h}}({\bf r}_{h})$,
and includes different three-particle $2e$--$h$ states
such that the total angular momentum projection
$M_z= N_e - N_h -m_1 -m_2 + M_h$ (and $N_e=n_1+n_2$, $N_h$) are fixed.
Here $\phi^{(e,h)}_{n m}$ are the $e$- and $h$- single-particle
factored wave functions in $B$; $n$ is the LL quantum number;
and $m$ is the oscillator quantum number [$m_{ze(h)}= {+ \atop (-)}(n-m)$].
We use the electron relative and center-of-mass coordinates:
${\bf r} = ({\bf r}_{e1} - {\bf r}_{e2})/\sqrt{2}$ and
${\bf R} = ({\bf r}_{e1} + {\bf r}_{e2})/\sqrt{2}$.
Permutational symmetry requires that for electrons
in the spin-singlet $s$ (triplet $t$) state
the relative motion angular momentum $n_1-m_1$ should be even (odd).
This basis complies with
the axial symmetry about the $z$-axis and the permutational symmetry.
The symmetry associated with magnetic translations
(e.g., \cite{Simon}) and its consequences are still hidden
at this point.

\begin{figure}[htb]
\begin{center}
\leavevmode
\epsfig{figure=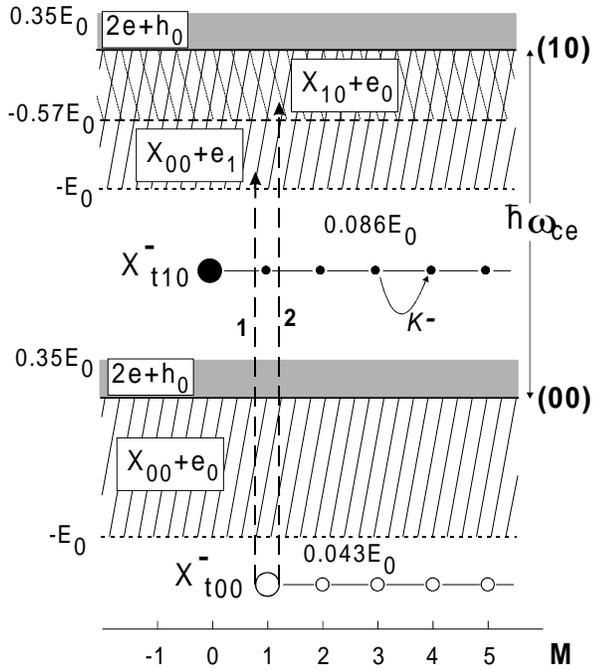,width=8cm}
\end{center}
\caption{
Schematic drawing of bound and scattering electron triplet $2e$--$h$
states
in the lowest LL's ($N_eN_h$)=(00), (10).
The quantum number $M=-M_z$ for the ($N_eN_h$)=(00) states
and $M=1-M_z$ for the ($N_eN_h$)=(10) states;
allowed transitions must satisfy $\Delta N_e = 1$; $\Delta M_z = 1$;
$\Delta k = 0$.
The energy $E_0= \protect\sqrt{\pi/2} \, e^2/\epsilon l_B$
parametrizes the 2D system in the limit of high $B$.
Large (small) dots correspond to the bound parent (daughter) $X^-$
states; see text for further explanations.}
\label{leveldiag}
\end{figure}

The calculated three-particle $2e$--$h$ eigenspectra
(electrons in the triplet state) in the two lowest LL's are shown
in Fig.\,\ref{leveldiag}.
Generally, the eigenspectra  associated with each LL
consist of bands of {\em finite} width $\sim E_0$.
The states within each such band form a {\em continuum}
corresponding to the extended motion of a neutral
magnetoexciton (MX) as a whole with the second electron
in a scattering state.
As an example, the continuum in the lowest ($N_eN_h$)=(00) LL
consists of the MX band of width $E_0$ extending down in energy
from the free $(00)$ LL. This corresponds to the $1s$ exciton ($N_e=N_h=0$)
plus a scattered electron in the zero LL, labeled $X_{00}+e_0$.
The structure of the continuum in the ($N_eN_h$)=(10) LL is
more complicated:
in addition to the $X_{00}+e_1$ band of the width $E_0$,
there is another MX band of width $0.574 E_0$ also extending down
in energy from the free ($N_eN_h$)=(10) LL. This
corresponds to the $2p^+$ exciton ($N_e=1$, $N_h=0$) \cite{L&L80}
plus a scattered electron in the $N_e=0$ LL, labeled $X_{10}+e_0$.
Moreover, there is a band above each free LL originating from the bound
internal motion of two electrons with the hole in a scattering
state (labeled $2e+h_0$) \cite{Dz&S_pr}.

Bound $X^-$ states (finite internal motions of all three particles)
lie {\em outside} the continua (Fig.\,\ref{leveldiag}).
In the limit of high $B$ the only bound $X^-$ state in the zeroth LL
($N_eN_h$)=(00) is the $X^-$-triplet. There are no bound
$X^-$-singlet states \cite{AHM,Whit97}
in contrast to the $B = 0$ case.
The $X^-$-triplet binding energy in zero LL's
($N_eN_h$)=(00) is $0.043 E_0$ \cite{AHM,Whit97}.
In the next electron LL ($N_eN_h$)=(10)
there are no bound $X^-$-singlets, and only one
bound triplet state $X^-_{t10}$, lying below the lower edge of the MX band
\cite{Dz&S_pr}.
The $X^-_{t10}$ binding energy is $0.086 E_0$, twice that of the $X^-_{t00}$,
and similar to the stronger binding of the $D^-$-triplet
in the $N_e=1$ LL \cite{Dz_PLA}.

We focus here on internal transitions in the $\sigma^+$ polarization
governed by the usual selection rules:
spin conserved, $\Delta M_z= 1$.
In this case the $e$-CR--like inter-LL ($\Delta N_e = 1$)
transitions are strong and gain strength with $B$.
Both
bound-to-bound $X^-_{t00} \rightarrow X^-_{t10}$
and photoionizing $X^-$ transitions are possible.
For the latter the final three-particle states in the $(10)$ LL belong
to the continuum (Fig.\,\ref{leveldiag}),
and calculations show that
the FIR absorption spectra reflect its rich structure \cite{Dz&S_pr}.
Transitions to the $X_{00}+e_1$ continuum are
dominated by a sharp onset at the edge (transition~1)
at an energy $\hbar\omega_{ce}$ plus
the $X^-_{t00}$ binding energy.
In addition, there is a broader and weaker peak corresponding to
the transition to the $X_{01}+e_0$ MX band, transition~2.
The latter may be thought of as the $1s\rightarrow 2p^+$
internal transition of the MX \cite{Int_X,Dz97},
which is shifted and broadened by the presence of the second electron.
Photoionizing transitions to the $2e+h_0$ band have extremely small
oscillator strengths and are not considered further.

The inter-LL bound-to-bound transition,
$X^-_{t00} \rightarrow X^-_{t10}$,
lies {\it below} the $e$-CR energy $\hbar\omega_{ce}$.
In contrast to the analogous, strong
triplet $T^-$ transition for $D^-$ \cite{Dz_PLA,Dz_D,Int_D}
this transition for $X^-$
has exactly {\em zero} oscillator strength,
a hidden manifestation of
the magnetic translational invariance \cite{Dz&S_pr}.
Consider the operator
${\bf \hat{K}} =
\sum_{j} (\bbox{ \pi}_j - \frac{e_j}{c} {\bf r}_j \times {\bf B})$,
which commutes with the Hamiltonian of interacting charged particles
in a uniform ${\bf B}$; here
$\bbox{ \pi}_j = -i \hbar \bbox{ \nabla}_j - \frac{e_j}{c} {\bf A}({\bf r}_j)$
is the kinematic momentum of the $j$-th particle.
The components of ${\bf \hat{K}} = (\hat{K}_x, \hat{K}_y)$
are the generators of magnetic translations
(\cite{Simon} and references therein)
and commute as canonically conjugate operators:
$[\hat{K}_x, \hat{K}_y] = i \frac{\hbar B}{c} \sum_j e_j$.
For the $X^-$, therefore, the operators
$\hat{k}_{\pm} = (\hat{K}_x \pm i \hat{K}_y) l_B/\sqrt{2}\hbar$
are the intra-LL lowering and raising operators.
Thus $(\hbar/l_B)^2 \hat{\bf K}^2$
has the oscillator eigenvalues $2k+1$ ($k=0, 1, \ldots$).
There is the macroscopic Landau degeneracy in the discrete quantum
number $k$, and
$k$ can be used, together with $M_z$,
to label the exact eigenstates.
The additional selection rule is conservation of
$k$ for dipole-allowed transitions \cite{footnote}.

As a result of the macroscopic degeneracy of $k$,
there exist {\em families} of bound,
degenerate $X^-$ states in $B$.
One family of $X^-$ bound states is associated
with the (00) and  (10) LL's.
Each $i$-th family starts with its {\it Parent State} (PS)
$|\Psi^{(P_i)}_{M_z}\rangle$, which has
the maximal possible value of $M_z$ in the family, (Fig.~\ref{leveldiag}).
The daughter states in the $i$-th family,
$|\Psi^{(D_i)}_{M_z-l}\rangle =
            \hat{k}_{-}^l |\Psi^{(P_i)}_{M_z} \rangle / \sqrt{l!}$
are constructed out of the PS
with the help of the raising operator $\hat{k}_{-}$ \cite{Dz&S_pr}.
Any PS carries the exact quantum number $k=0$
(while a daughter state in the $l$-th generation carries $k=l$).
The additional requirement of conservation of $k$ for allowed
electric-dipole transitions means that {\em parent states},
$|\Psi^{(P_i)}_{M_z}\rangle$ and $|\Psi^{(P_j)}_{M_z'}\rangle $
in the two families,
must satisfy
$M_{z}' = M_{z} \pm 1$
for there to be allowed FIR transitions. The PS's in the (00) and (10)
LL's have
$M_{z} = -1$ and
$M_{z}' = +1$,
and do not satisfy this requirement.
Therefore, the family of $X^-_{t10}$ bound states is {\em dark},
i.e., is not accessible by internal transitions from the ground
$X^-_{t00}$ bound states. Breaking of translational invariance
(by impurities, disorder etc.) would make the
$X^-_{t00} \rightarrow X^-_{t10}$ transition allowed.

\begin{figure}[htb]
\begin{center}
\leavevmode
\epsfig{figure=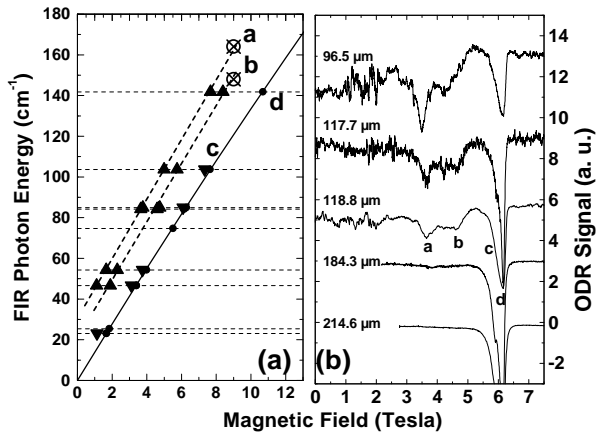,width=8cm}
\end{center}
%
\caption{{\bf (a)} Summary plot of the observed internal
transitions of \xminus
as a function of magnetic field. Dashed lines are guides to the eye,
the solid line represents a fit through the measured positions of
electron cyclotron resonance. The crossed out circles at $9\,T$ are
the results of the calculation described in the text.
{\bf (b)} ODR scans as a function of magnetic field for
several FIR laser wavelengths.
The horizontal (B) scale corresponds to the scan taken at 118.8\,$\mu
m$.
(a) and (b) denote the \xminus-singlet transitions and (c)
marks the \xminus-triplet transition.}
\label{summaryplot}
\end{figure}

Most qualitative features discussed above are preserved at finite
fields and confinement
where both triplet and
singlet bound $X^-$ states exist, as shown
by calculations for a representative case
(200\,\AA\ GaAs/Ga$_{0.7}$Al$_{0.3}$As QW at $B > 9$\,T).
We obtain eigenstates of $X^-$ using an
expansion \cite{Dz_D,Whit97}  in $e$  and $h$
LL's and size-quantization levels in a QW;
we assume a simple valence band for holes with
in-plane  $m_{h\|}=0.24$ and perpendicular $m_{hz}=0.34$
effective masses ($m_{e}=0.067$ is isotropic, the GaAs dielectric
constant $\epsilon = 12.5$).
For the singlet and triplet binding energies of $X^-$ in the zeroth
LL
we obtain results equivalent to the high-accuracy calculations
of \cite{Whit97}.
Both singlet and triplet transitions
exhibit a characteristic double-peak structure,
but the singlet transitions are broader,
and the peak at higher
energies has a larger oscillator strength. Results for the singlet
transitions at 9 Tesla are shown in Fig.~\ref{summaryplot}(a).
%
%
%
For comparison with the theoretical calculations
we have studied two 20\,nm wide GaAs/Al$_{0.3}$Ga$_{0.7}$As
multiple quantum well (MQW) structures, one undoped and one
modulation-doped in the barrier with silicon donors at
$2\times10^{10}$\,cm$^{-2}$, by optically detected resonance
(ODR) spectroscopy \cite{odrtechnique}.
Both samples show the
\xminus-photoluminescence (PL) line.
The experiment confirms the theoretical predictions discussed above.
In Fig.~\ref{summaryplot}(b) a series of ODR scans for the 
undoped sample is shown at
several FIR laser wavelengths. 
During the magnetic field sweep the spectrometer was
stepped to remain centered on the peak 
of the \xminus recombination line. The recorded changes in the PL
intensity ($\Delta I_{\mbox{PL}}$) of the \xminus-PL line 
correspond to a decrease in the PL intensity of \xminus. 
Tracking the PL peak of X instead of \xminus yields ODR scans which are
inverted (increase in PL strength), but otherwise very similar. 
The spectra in Fig.~\ref{summaryplot}(b) 
are displaced in magnetic field
such as to align the position of the sharp, negative going
electron cyclotron resonance (e-CR) 
present in all scans at the position of 
the e-CR at 118.8\,$\mu m$, 6.23 Tesla. 
In this way, the behavior of
features occurring at lower magnetic fields, and with
an amplitude approximately one order of magnitude smaller than e-CR,
are more easily seen.

These ODR scans show several resonances at magnetic fields below
that of e-CR. For instance, the ODR scan recorded at a FIR wavelength
of 118.8\,$\mu m$ has features at 4.08 Tesla (a) and 4.60 Tesla (b), 
in addition
to a shoulder roughly 0.1 Tesla (c) below e-CR (d). Resonances 
of similar shape
are observed in all ODR scans recorded with FIR
wavelengths shorter than 393.6\,$\mu m$ . At 393.6\,$\mu m$
and 432.6\,$\mu m$ no
resonances were observed at magnetic fields below e-CR.
The qualitative behavior of the 
modulation-doped sample is very similar;
however, the strength of features (a) and (c) at lower
magnetic fields is greatly enhanced with respect to e-CR. In this
sample the 
intensity of the \xminus-PL line is
much larger than in the undoped structure, so \xminus-features are
expected to be stronger.

The observed resonances are summarized in
Fig.~\ref{summaryplot}(a), where 
the peak positions in the ODR scans for the undoped sample
are plotted as a function of magnetic field
for all measured FIR laser lines. Features labeled (a) and (b) in
Fig.~\ref{summaryplot}(b) are shown by upright triangles, feature (c) is
represented by an inverted triangle, and e-CR is marked by solid circles.
Features (a) and (c) occur in the modulation-doped sample
approximately 0.1 Tesla lower than the data points in this figure.
Feature (b) is observed at elevated temperature.
Two points from the numerical calculation for the singlet
transitions at 9
Tesla are plotted as crossed-out 
circles at the end of the two dashed lines, which are guides to the eye. 
The predicted double-peak structure 
for the \xminus singlet transition is clearly 
observed and in good agreement with the calculations. 

Based on the discussion above
and the good qualitative as well as quantitative agreement of the
experiment with theory, we assign features (a) and (b) to the
internal \xminus-singlet transition, and feature (c) to the internal
\xminus-triplet transition.
No evidence was found for any features 
on the high-field size of e-CR corresponding to the
localized-to-localized state transitions
({\em e.g.} the \xminus-triplet) that are 
dominant in magneto-spectroscopy of \dminus. 
%
%
%
In conclusion, photoionizing spin-singlet and spin-triplet
transitions of $X^-$ have been observed experimentally.
The experimental findings are in good agreement
with theoretical predictions,
which show that due to an additional exact selection rule
resulting from a hidden symmetry,
the bound-to-bound $X^-$ singlet- and
triplet-transitions to the next electron LL
are forbidden. The appearance of a corresponding triplet feature
{\em below} the $e$-CR energy should be a characteristic feature associated
with breaking of translational invariance. This may be
used as a tool for studying the extent of $X^-$ localization.

ABD is grateful to the Department of Physics, University at Buffalo,
where part of this work was performed, and 
to the Humboldt Stiftung for financial support.
The work at SUNY at Buffalo was supported by NSF grant DMR 9722625.
%
%

%

\end{multicols}

\begin{references}
\bibitem{footnote}
The Hamiltonian of the interaction with the
FIR radiation of polarization $\sigma^{\pm}$ is of the form
$\hat{V}^{\pm} \sim  \sum_{j} e_j \pi_{j}^{\pm}/m_j$,
where $\pi_j^{\pm} = \pi_{jx} \pm i \pi_{jy}$, and
commutes with $\hat{\bf K}$ (and $\hat{\bf K}^2$).
\end{references}
\end{document}